\documentclass[twocolumn,twoside,slac_two]{revtex4}
\usepackage{graphicx}
\usepackage{amsmath}
\usepackage{fancyhdr}
\newcommand{\beq}{\begin{equation}}
\newcommand{\eeq}{\end{equation}}

\begin{document}

\title{Predictions for $\sin 2(\beta/\phi_1)_{\rm eff}$ in $b\to s$ penguin dominated modes}

%

\author{J. Zupan}
\affiliation{Department of Physics, University of Ljubljana, Jadranska 19, 1000
Ljubljana, Slovenia}
\affiliation{J. Stefan Institute, Jamova 39,
P.O. Box 3000, 1001 Ljubljana, Slovenia}

\begin{abstract}
We provide a review of predictions for $\sin 2\beta_{\rm eff}$ in $b\to s$ penguin dominated modes based on $1/m_b$ expansion and/or SU(3) flavor symmetry. 
The experimental results are consistently lower than the theoretical predictions. In order to interpret whether this effect is a sign of
new physics contributions or can be explained away within the Standard Model a theoretical input cannot be avoided. The effect survives at a level larger than $2.1 \sigma$ in a conservative average over different modes that includes theoretical predictions.
\end{abstract}

\maketitle

\thispagestyle{fancy}


\section{Introduction}
A nontrivial test of the Standard Model (SM) are the two ways of measuring $\sin 2\beta$ from time dependent $\Delta S=1$ $B$ decays [with $\beta=\arg(-V_{cd}V^*_{cb}/V_{td}V^*_{tb})$]: (i) from tree dominated, e.g. $B\to J/\Psi K_S$ \cite{Bigi:1981qs}, and (ii) from penguin dominated, e.g. $B\to\phi K_S$, decay modes \cite{London:1989ph,Grossman:1996ke}. The
two determinations should be the same in the SM, but would differ, if new physics contributions modify
the penguin dominated decay amplitudes. For several years now there is some disagreement
between the two determinations, if the CKM suppressed terms are neglected in the interpretation of the experimental results. However, with the decreased experimental errors this approximation is no more adequate. As I will argue in this write-up theoretical input is needed for the correct interpretation of experimental results.

The two observables measured in time dependent $B(t)\to f$ decays into a CP eigenstate $f$  are the indirect CP asymmetry
\beq
S_f= 2 \frac{\rm{Im}\big[ e^{-i 2\beta} \bar A_{f}/{A_{f}}\big]}{1+ |\bar A_{f}|^2/|{A_{f}}|^2}, \label{Sf}
\eeq
and the direct CP asymmetry
\beq
C_f=\frac{|{A_{f}}|^2- |\bar A_{f}|^2}{|{A_{f}}|^2+ |\bar A_{f}|^2}.
\eeq
Above we have used the notation for the decay amplitudes $A(\bar B^0\to f)=\bar A_f$ and $A(B^0\to f)=A_f$. The choice of  $\Delta S=1$ $B^0$ decays makes
the determination of $\sin 2\beta$ from $S_f$ theoretically very clean since it exploits the CKM hierarchy $\lambda_u=V_{ub}V_{us}^*\sim \lambda^2 \lambda_c$, where $\lambda_c=V_{cb}V_{cs}^*$ and $\lambda=\sin \theta _C=0.22$. To see this let us 
split the amplitude according to the CKM factors
\beq
\begin{split}
\bar A_f&=\lambda_c a_f^c+\lambda_u a_f^u+\lambda_t a_f^t\\
&=\lambda_c (a_f^c-a_f^t)+\lambda_u (a_f^u-a_f^t)\\
&=\lambda_c A_f^c+\lambda_u A_f^u,\label{Afdecomp}
\end{split}
\eeq
where in obtaining the second row the CKM unitarity $\lambda_c+\lambda_u+\lambda_t=0$ was used. The different terms in Eq. \eqref{Afdecomp} can receive the following contributions, depending on the final state $f$: $a_f^c$ can receive contributions from $b\to c\bar c s$ tree and $c\bar c$ rescattering (charming penguin); $a_f^u$ can receive contributions from $b\to u\bar u s$ tree and $u\bar u$ rescattering ($u-$penguin); $a_f^t$ can receive contributions from QCD penguins and electroweak penguins. 

Since $\lambda_u\sim 0.02\lambda_c$ there is a big hierarchy between the two terms in $\bar A_f=\lambda_c A_f^c+\lambda_u A_f^u$, so that $\bar A_f$ is dominated by one CKM amplitude. Since $\lambda_c$ is real in the standard CKM parametrisation, $\bar A_f\simeq A_f$, and the ratio of the two amplitudes cancels to first approximation in  Eq. \eqref{Sf}. More precisely, expanding in the small ratio
\beq
r_f e^{i\delta_f}=\left|{\lambda_u}/{\lambda_c}\right|\cdot \frac{A_f^u}{A_f^c}\simeq 0.02 \frac{A_f^u}{A_f^c}, \label{rf}
\eeq
we have 
\beq
\begin{split}
\sin 2 \beta_{\rm eff}&\equiv-\eta_f^{CP} S_f=\\
&=\sin 2 \beta + 2  r_f \cos \delta_f \cos 2 \beta \sin \gamma , \label{betaeff}
\end{split}
\eeq
where $\eta_f^{CP}$ is the CP of the final state $f$, and 
\beq
C_f=-2r_f \sin \delta_f \sin \gamma .  
\eeq
If the small $r_f$ terms are neglected we thus have  $\sin 2 \beta_{\rm eff}=\sin 2 \beta$ and $C_f=0$. If a nonzero direct CP asymmetry $C_f$ is found experimentally, it would immediately imply that $r_f$ terms are important. 


\begin{figure}
\begin{center}
\includegraphics[width=8.cm]{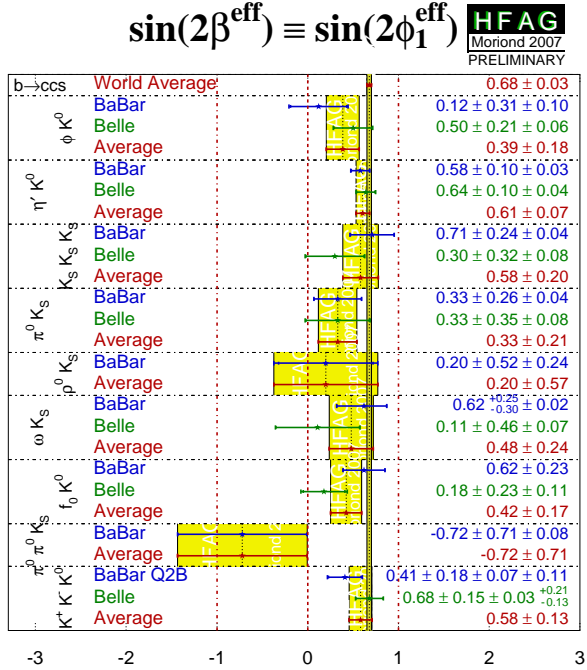}
\end{center}
\caption{The measured $\sin 2\beta_{\rm eff}^{\rm Peng}$ for $\Delta S=1$ penguin dominated decays \cite{Barberio:2007cr}. The two vertical yellow lines
give the $\sin 2\beta_{\rm eff}^{\rm Tree}$ from $b\to c\bar c s$ world average. } \label{Exp:res}
\end{figure}

\section{Two ways to $\sin 2\beta$}
As alluded to in the introduction, it is useful to distinguish two determinations of $\sin 2\beta$. The tree dominated decays, e.g. $B^0\to J/\Psi K_S$, are 
expected to be SM dominated. We will denote the corresponding value in Eq. \eqref{betaeff} as  $\sin 2\beta_{\rm eff}^{\rm Tree}$. 
The penguin dominated decays, e.g. $B^0\to \phi K_S$, can on the contrary receive
possibly large beyond Standard Model contributions. The corresponding values in Eq. \eqref{betaeff} will be denoted as $\sin 2\beta_{\rm eff}^{\rm Peng}$. The comparison of the two then tests the KM mechanism
\beq
\begin{split}
\Delta S_f&=\sin 2\beta_{\rm eff}^{\rm Peng}-\sin 2\beta_{\rm eff}^{\rm Tree}\\
&=O(r_f^{\rm Peng})-O(r_f^{\rm Tree}).
\end{split}
\eeq
The $O(r_f^{\rm Tree})$ difference between $\sin 2\beta$ and $S_{J/\Psi K_S}$ is below a percent level, since $A_f^u$ in Eq. \eqref{rf} is already at least $\alpha_S(m_b)$ suppressed compared to the dominant tree term, $A_f^c$ \cite{Gronau:1989ia,Boos:2004xp,Ciuchini:2005mg,Li:2006vq}. These $O(r_f^{\rm Tree})$ corrections will be neglected compared to the $O(r_f^{\rm Peng})$ differences between $\sin 2\beta_{\rm eff}^{\rm Peng}$ and $\sin 2 \beta$ which we will investigate below. 

The expected difference $\Delta S_f$ for penguin dominated modes is channel dependent. Curiously enough, the experimental values are all negative, $\Delta S_f<0$, see Fig. \ref{Exp:res}.  This experimental pattern immediately raises several questions
\begin{itemize}
\item
what are the SM expectations?
\item
what are the errors on the theory predictions?
\item
what theoretical errors to expect in the future/can we improve them?
\end{itemize}
The last question is especially interesting for future prospects, where with 50~ab${}^{-1}$ of data $S_{\phi K_S}$ and $S_{\eta' K_S}$ are expected to be measured to a precision of a few 
percent. 

\begin{figure}
\begin{center}
\includegraphics[width=7.5cm]{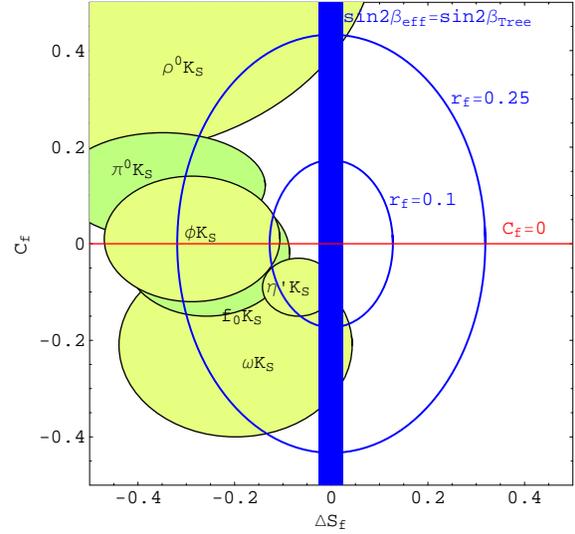}
\end{center}
\caption{The $1\sigma$ experimental values for ($\Delta S_f, C_f$) in penguin dominated modes as of FPCP07 conference \cite{Barberio:2007cr}. The vertical
blue band shows experimental errors on $\sin 2\beta_{\rm eff}^{\rm Tree}$ from $b\to c\bar c s$ modes. The two blue circles represent $\Delta S_f$ for $r_f=0.1, 0.25$ with $\delta_f$ varied (with $\gamma =60^\circ$).} \label{fig-ellipses}
\end{figure}

An important thing to note is that we have 2 observables,  $S_f$ and $C_f$, but also 2 unknowns: $\sin\gamma~r_f  $ and  $\delta_f$
\begin{align}
\Delta S_f&=2  \sin\gamma ~r_f \cos \delta_f\cos 2 \beta  \\
C_f&=-2\sin\gamma ~ r_f \sin \delta_f  
\end{align}
To predict $\Delta S_f$ one therefore {\it necessarily} needs theory input at least on $r_f$, while $\delta_f$ could in principle 
be fixed from a measurement of $C_f$ (or vice versa). An example of this is shown in Fig. \ref{fig-ellipses}, where the experimental results are compared with ellipses in 
$(\Delta S_f, C_f)$ plane obtained for $r_f=0.1,0.25$ and arbitrary $\delta_f$ (and with $\gamma$ chosen to be $60^\circ$). Note that these two values of $r_f$ correspond to fairly large 
values of $A_f^u/A_f^c\sim 5, 10$ in Eq. \eqref{rf}.

Both $\Delta S_f$ and $C_f$ have been estimated in several theoretical frameworks using SU(3) flavor symmetry and using $1/m_b$ expansion: QCDF, SCET, pQCD. We
discuss these two approaches next.

\section{Using flavor SU(3)}
\begin{figure}
\includegraphics[width=3.1cm]{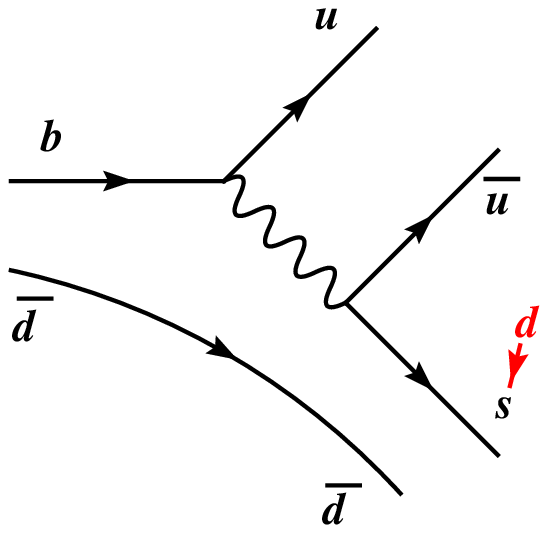}
~~~~~\includegraphics[width=4.1cm]{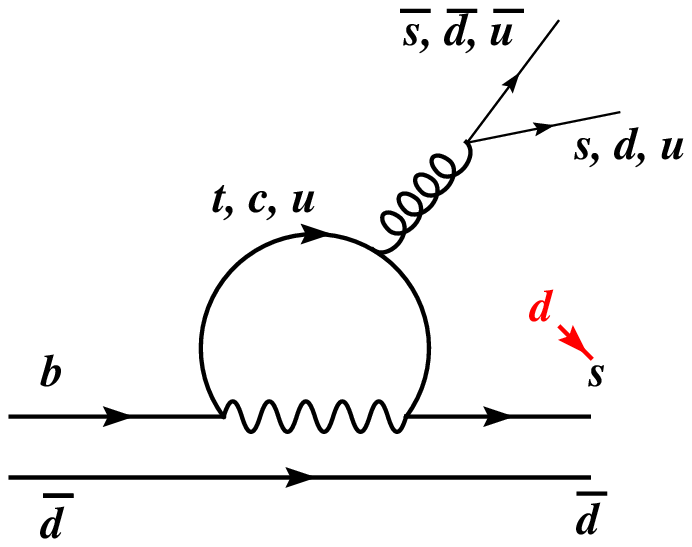}
\caption{The $s\to d$ exchange modifies the hierarchy of tree (left diagram) and penguin (right diagram) contributions by replacing $V_{ts}\to V_{td}$ and  $V_{us}\to V_{ud}$ respectively.}\label{SU3Fig}
\end{figure}

As pointed out in \cite{Grossman:2003qp} and discussed later also in \cite{Gronau:2004hp,Gronau:2003kx, Gronau:2006qh,Gronau:2005gz,Raz:2005hu,Engelhard:2005ky,Engelhard:2005hu}
one can use $\Delta S=0$ modes related by $SU(3)_F$ (represented by $s\to d$ exchange on Fig. \ref{SU3Fig}) to constrain $\Delta S_f$ in penguin dominated $\Delta S=1$ 
decays. This corresponds to a replacement $V_{cb}V_{cs}^*A_f^c\to V_{cb}V_{cd}^*A_f^c{}'$ and 
$V_{ub}V_{us}^*A_f^u\to V_{ub}V_{ud}^*A_f^u{}'$ in Eq. \eqref{Afdecomp}, where the primes remind us of the fact that one needs to take into account SU(3) breaking as well as of the fact that $f$ may transform into a sum of mass eigenstates (for instance U-spin transforms $\pi^0\sim (u\bar u- d \bar d)/\sqrt{2}$ to $(u\bar u- s \bar s)/\sqrt{2}$, which is a sum of $\eta$ and $\eta'$).

In the SU(3) related amplitudes the hierarchy of tree and penguin contributions is changed because the CKM factors in front of the matrix elements $A_f^{c,u}$ in Eq. \eqref{Afdecomp} have changed
\beq
P\to -\lambda P',\quad T\to T'/\lambda.
\eeq
For instance, the $B\to \pi K$ amplitudes are penguin dominated, while in SU(3) related $B\to \pi\pi$ decays the tree contributions are larger than the penguins.
Because of this, one  can bound "tree pollution" $r_f$ in $\Delta S=1$ decays from the related $\Delta S=0$ modes. A bound on $r_f$ consists of a sum over modes
\beq
r_f \le \frac{{\cal R} + \bar \lambda^2}{1 - {\cal R}}, \quad
{\cal R} \le \bar \lambda \sum_{f'} |a_{f'}| \sqrt{\frac{\bar {\cal B}_{f'}(\Delta S=0)}{\bar {\cal B}_f
(\Delta S=1)}},\label{Rbound}
\eeq
where $a_{f'}$ are numerical coefficients. From the above equation we immediately see that the bound can never be better than $r_f< \bar \lambda^2\sim 0.05$, even if 
${\cal R}$ is set to zero. 

\begin{figure}
\begin{center}
\includegraphics[height=6.2cm]{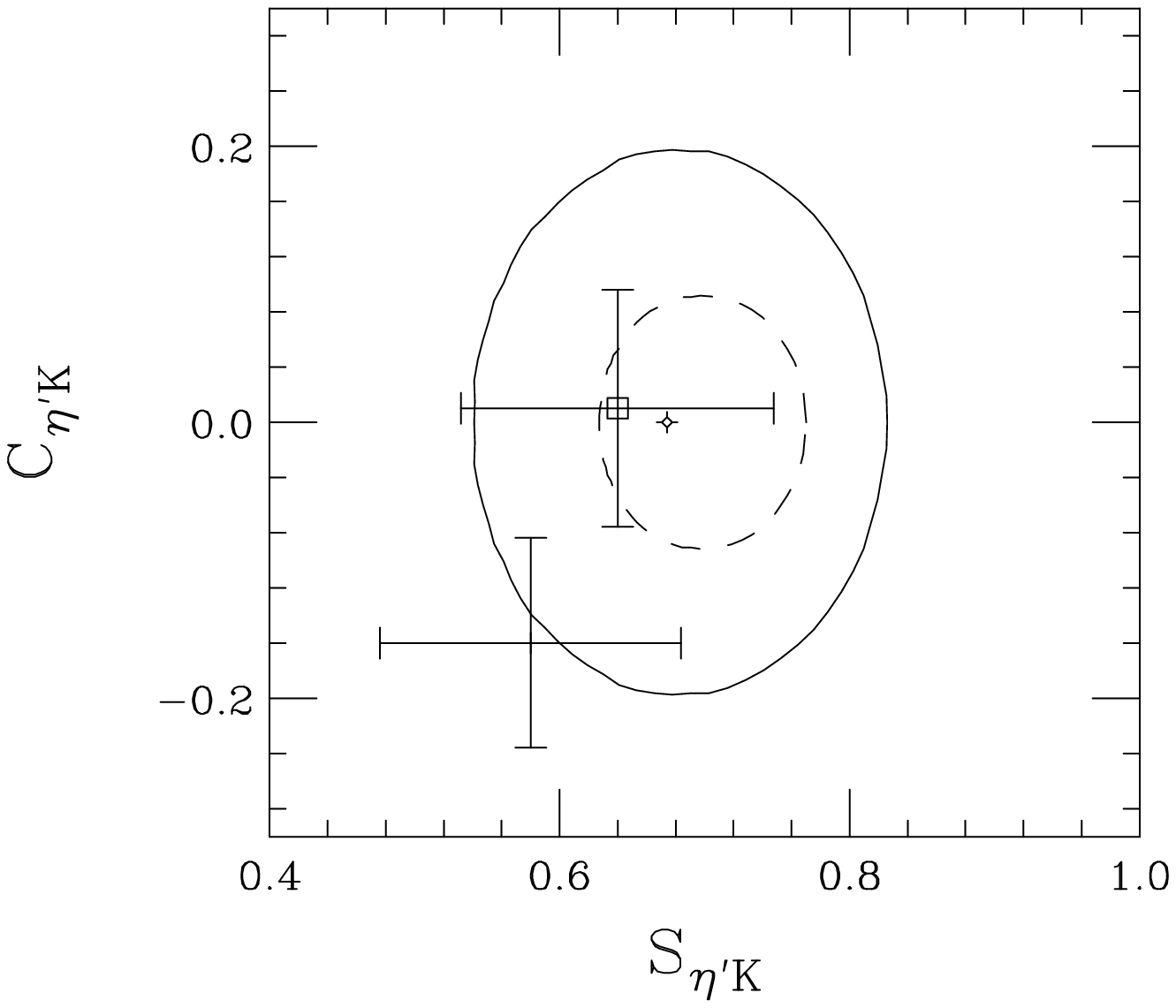}\\
${}$\\
\includegraphics[height=4.cm]{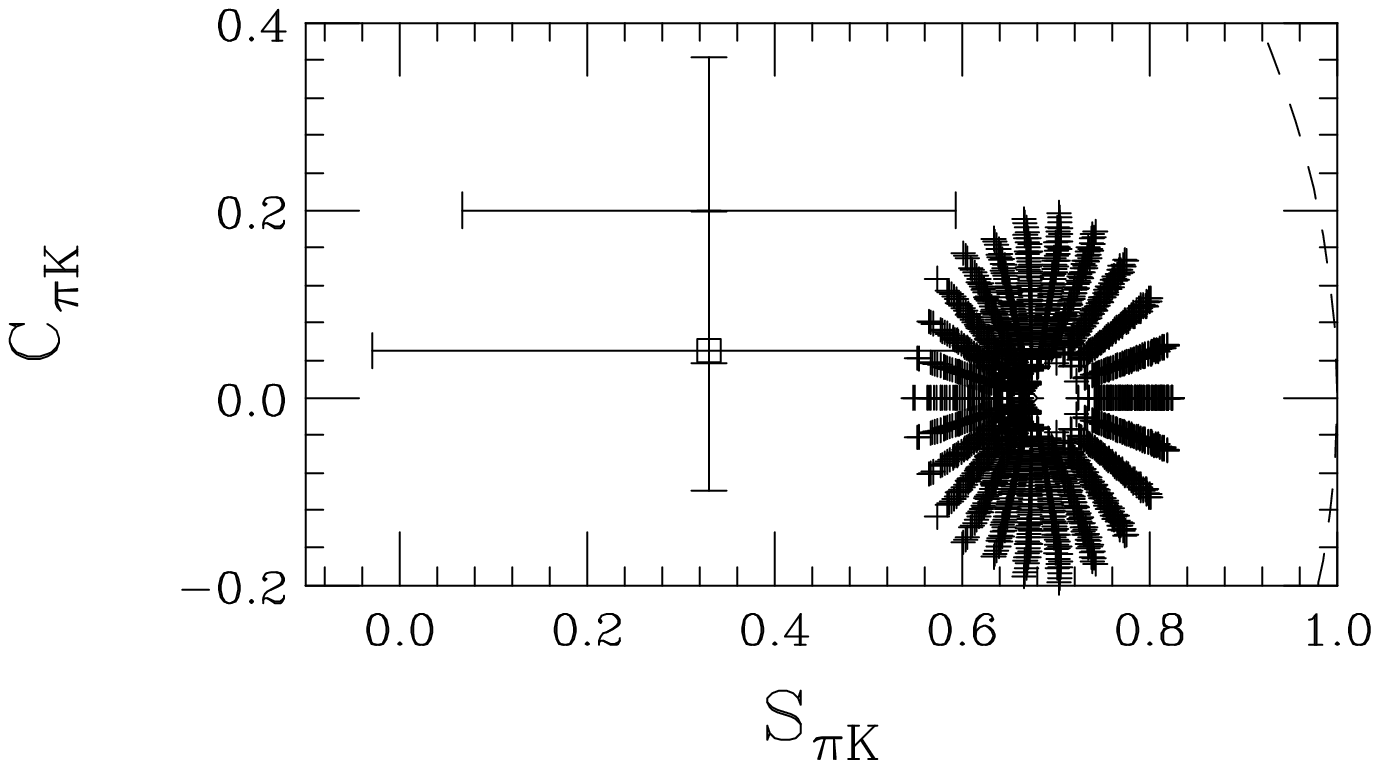}
\end{center}
\caption{Top: $(S_{\eta'K_S}, C_{\eta'K_S})$ values  allowed by SU(3) bounds (region enclosed by the solid curve), and with further dynamical assumptions (region enclosed by the dashed curve) \cite{Gronau:2006qh}. Bottom: $(S_{\pi^0 K_S}, C_{\pi^0K_S})$ values allowed by SU(3) bounds \cite{Gronau:2006qh}. The small points are  $(S_f, C_f)=(\sin 2\beta,0)$. Experimental values are from BaBar (dot) \cite{Aubert:2006wv,Aubert:2006ad} and from Belle (square) \cite{Chen:2006nk,Hara:Belle}. }\label{Fig:update}
\end{figure}

The upper bound on ${\cal R}$ in Eq. \eqref{Rbound} was obtained by bounding a sum over amplitudes, where there would be in general cancellations between different terms, with a sum over absolute values of amplitudes, where of course no such cancellations occur. The bound on ${\cal R}$ is thus in general better, if the sum is over a smaller set of modes $f'$. Furthermore, all the branching ratios $f'$ in the bound need to be measured to have the best bound. At present for some $\Delta S=0$ modes only upper bounds are known. For instance in the bound on $r_{\eta'K_S}$  the branching ratios for $B^0\to\pi^0\eta, \eta^{(')}\eta^{(')}$ decays enter. For these only experimental upper bounds exist, giving at present ${\cal R}_{\eta'K_S}<0.116$, while one arrives at ${\cal R}<0.045$, if the predicted branching ratio in QCDF, Scenario 4,  are used (or ${\cal R}<0.088$ if SCET, Sol. I., predictions are used). Clearly, there is still room for improvement using this approach. But in general, assuming only SU(3) without any dynamical assumptions, gives too conservative bounds. The reason is that in this way one does not use any information about the relative phases between the terms in 
the sum in Eq. \eqref{Rbound}. The results of a 2006 numerical update \cite{Gronau:2006qh}, where correlations between $S_f$ and $C_f$ were used, are shown on Fig. \ref{Fig:update}. Bounds on $\Delta S_{\phi K_S}$ are much worse \cite{Grossman:2003qp}. It is also possible to treat $S_{KKK}$ in this framework, however, the bounds are again not very informative \cite{Engelhard:2005ky,Engelhard:2005hu}. Assuming small annihilation one has 
$r_{K^+K^-K^0}<1.02,$ and  $r_{K_S K_S K_S}<0.31$ \cite{Engelhard:2005ky,Engelhard:2005hu}.

Another use of SU(3) is to perform global fits to the data \cite{Chiang:2003pm,Chiang:2004nm,Buras:2003dj,Fleischer:2007mq}. In this case $\Delta S_f$ can be predicted and not just bounded as above. A recent analysis in \cite{Fleischer:2007mq} shows a discrepancy between experimental
data and the SU(3) fit predictions in the $(S_{\pi^0K_S},C_{\pi^0K_S})$ plane. The fit predicts $S_{\pi^0K_S}= -0.81 \pm 0.03$ in the Standard Model, which is to be compared with the measured value of $S_{\pi^0K_S}=-0.33 \pm 0.21$. Note that the error on the prediction already includes the variation due to the SU(3) breaking.

\section{Using $1/m_b$ expansion}
The $1/m_b$ expansion has more predictive power. I would like to stress that $1/m_b$ expansion is a consistent framework, based on Soft Collinear Effective Theory \cite{Bauer:2000ew}. Like the SU(3) approach it is in principle "model independent" in the sense that it uses only symmetries of QCD. While the SU(3) approach uses a symmetry that arises in the $m_s\to 0$ limit, SCET based approaches use the symmetry that arise in the $m_b\to \infty$ limit. The framework offers consistency checks both within two-body $B$ decays as well as in $B\to D\pi$ \cite{Mantry:2003uz} and semiinclusive hadronic decays \cite{Chay:2006ve,Soni:2005jj}. Note that both QCD Factorization (QCDF) \cite{Beneke:1999br,Beneke:2002jn,Beneke:2003zv} and the so-called SCET calculations \cite{Bauer:2004tj,Jain:2007dy,Williamson:2006hb} use Soft Collinear Effective Theory, but they differ in the treatment of subleading effects and charming penguin contributions \cite{Bauer:2005wb,Beneke:2004bn}. 

We first review state of the art in these calculations and then move on to the predictions in specific decay modes. Both in 
QCDF and SCET the hard kernels are known to NLO in $\alpha_S(m_b)$  \cite{Beneke:1999br,Chay:2003ju,Jain:2007dy,Beneke:2005vv}, with partial results already known at NNLO \cite{Bell:2007tv}. The jet functions are known to NLO in  $\alpha_S(\sqrt{\Lambda m_b})$ \cite{Hill:2004if,Kirilin:2005xz,Beneke:2005gs}. At present the limit on accuracy is the inclusion of $1/m_b$ corrections. While some of them, for instance the chirally enhanced terms, have already been included \cite{Beneke:1999br,Jain:2007dy}, more work is needed to complete the calculations to $1/m_b$ order. 

Not all of this information was used in $\Delta S_f$ calculations, however. In most recent QCDF calculation of Ref. \cite{Beneke:2005pu} hard scattering was treated at LO in $\alpha_S(m_b)$, $\alpha_S(\sqrt{\Lambda m_b})$, soft overlap at NLO in $\alpha_S(m_b)$ and some $1/m_b$ corrections were included (modeled). In SCET calculation \cite{Williamson:2006hb} all hard kernels were taken at LO in $\alpha_S(m_b)$, jet functions were not expanded in $\alpha_S(\sqrt{\Lambda m_b})$, $1/m_b$ corrections were not included, while nonperturbative  parameters (also the charming penguin one, $P_{\rm charm}$) were fit from data. In pQCD calculations \cite{Li:2005kt,Li:2006jv} the soft overlap contribution is  factorized and some NLO corrections are included.

An interesting way of using the $1/m_b$ expansion results was proposed by M. Ciuchini {\it et al.} \cite{Ciuchini:2001gv,Silvestrini:2007yf}. Here the renormalization group invariant parametrization of the decay amplitudes \cite{Buras:1998ra} is used to fit from data the $1/m_b$ corrections to the QCDF predictions. In this way a better desription of branching ratios and CP asymmetries is obtained. The predictions on $\Delta S_f$ are compatible with the original QCDF predictions, albeit with larger errors \cite{Silvestrini:2007yf}. The errors will shrink once more data on relevant branching ratios and direct CP asymmetries become available.

\subsection{$\Delta S$ for $\phi K_S$}
\begin{figure}
\begin{center}
\includegraphics[width=7.8cm]{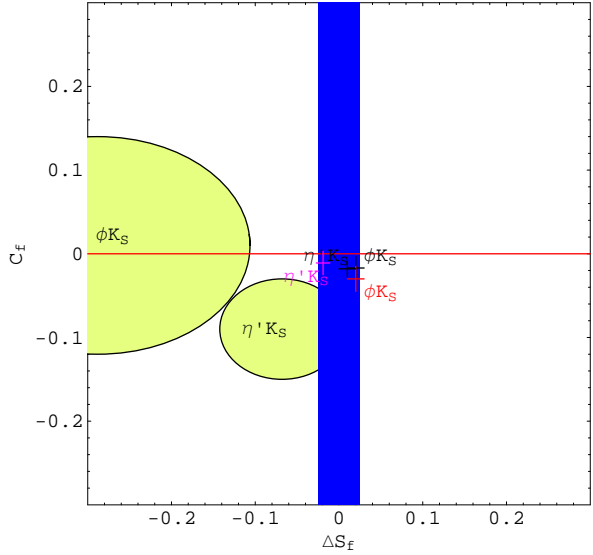}
\end{center}
\caption{Crosses: QCDF (black) \cite{Beneke:2003zv}, pQCD(red) \cite{Li:2006vq}, and SCET (magenta) \cite{Williamson:2006hb} predictions for $\Delta S_f, C_f$, with $f=\phi K_S, \eta' K_S$. Ellipses are experimental $1\sigma$ allowed regions, blue band is experimental error on $\sin 2\beta$ from $b\to c\bar c s$.}
\end{figure}

This is the cleanest mode, with the least ambiguity on $\Delta S_f$, since there is no $b\to u\bar u s$ tree contribution. One thus has
\beq
\frac{A_f^u}{A_f^c}=\frac{a_f^u-a_f^t}{a_f^c-a_f^t}\sim O(1), \label{phiKSratio}
\eeq
where $a_f^i$ are either $\alpha_S(m_b)$ (penguins) or $1/m_b$ suppressed. 
The "tree pollution" parameter $r_f$ is then at a percent level as demanded by the CKM suppression,
$r_f\simeq 0.02 {A_f^u}/{A_f^c}$. In particular, the ratio of the matrix elements, ${A_f^u}/{A_f^c}$, cannot be enhanced, since there is no tree
contribution to ${A_f^u}$. In accordance with this
expectation both calculations in QCDF \cite{Beneke:2005gs} and pQCD \cite{Li:2006jv} obtain
\beq
\Delta S_{\phi K_S}=0.02\pm0.01,
\eeq
while there is no  prediction in SCET yet. An analysis in \cite{Cheng:2005bg} suggest that final state interactions do not change the above result. 

\subsection{$\Delta S$ for $\eta' K_S$}
Because $\eta'$ contains a $u \bar u$ component there is a $b\to u\bar u s$ tree level contribution to the 
$B\to \eta K_S$ decay amplitude. However, $r_f$ is still small, since $A_f^c$ is also enhanced. This enhanced $A_f^c$ explains the 
large $Br(B\to \eta'K_S)$ observed experimentally. The enhancement itself can be understood through constructive interference between $A(B\to \eta_{q}K_S)$ and $A(B\to \eta_{s}K_S)$,
a mechanism that also explains small   $Br(B\to \eta K_S)$, where the interference is destructive \cite{Lipkin:1990us,Lipkin:1998ew}. Besides the interference pattern gluonic contributions and/or 
SU(3) breaking are needed  to
obtain the experimentally observed branching ratios \cite{Beneke:2002jn,Beneke:2003zv,Gerard:2006ch,Williamson:2006hb}. 

The nonperturbative
parameters including gluonic charming penguins were fit from experimental data in SCET \cite{Williamson:2006hb} (but not from $\Delta S_{\eta'K_S}$, which is a pure prediction), while in QCDF calculation of \cite{Beneke:2003zv} a reasonable estimate for these unknown terms was used. The two predictions
\beq
\begin{split}
{\rm QCDF:~} \Delta S_{\eta'K_S}=&0.01\pm0.01,\\
{\rm SCET:~} \Delta S_{\eta'K_S}=&
\left\{
\begin{matrix}
-0.019\pm0.008, {\rm ~Sol.~I},\\
-0.010\pm0.010, {\rm ~Sol.~II},
\end{matrix}
\right. 
\end{split}
\eeq 
do not coincide, but both of them do consistently give small deviations. This would be true also if, for some reason, the strong phases between $A^c$ and $A^u$ were 
completely missed in the calculation, since $|A^u/A^c|$ is $O(1)$ as in $B\to \phi K_S$, Eq.  \eqref{phiKSratio}, and is not enhanced despite the presence of a tree contribution. 
The situation is reversed in $B\to \eta K_S$, where the destructive interference between $A(B\to \eta_{q}K_S)$ and $A(B\to \eta_{s}K_S)$ suppresses $A_f^c$ and makes the tree contribution relatively larger. Then $\Delta S_{\eta K_S}$ can be large, even $O(1)$.

\subsection{Other 2-body modes}
\begin{figure}
\begin{center}
\includegraphics[width=7.8cm]{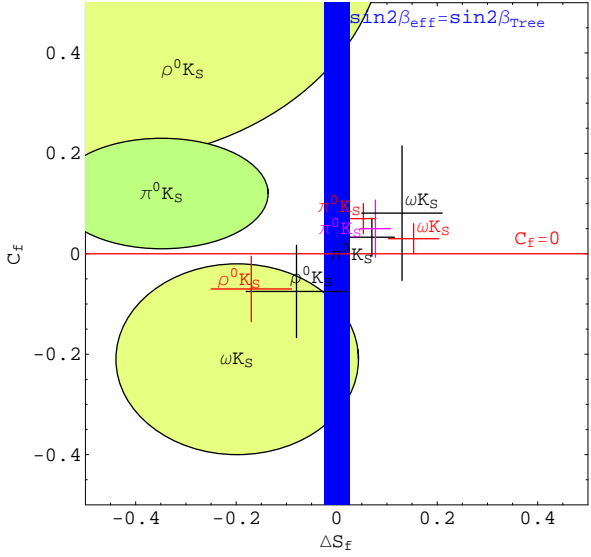}
\end{center}
\caption{Crosses: QCDF (black) \cite{Beneke:2003zv}, pQCD(red) \cite{Li:2006vq,Li:2005kt}, and SCET (magenta) \cite{Williamson:2006hb} predictions for $\Delta S_f, C_f$. Ellipses are experimental $1\sigma$ allowed regions, blue band is experimental error on $\sin 2\beta$ from $b\to c\bar c s$.} \label{fig:other2-body}
\end{figure}

The other 2-body modes for which there exist predictions on $\Delta S_f$ are  ${\pi^0 K_S}$, ${\rho^0 K_S}$ and ${\omega K_S}$. All of these receive $b\to u\bar u s$ tree contributions, so that $A^u$ is enhanced over $A^c$. In general one expects $\Delta S_f$ $\sim O(0.1)$, with calculated values given below
{\tiny
\begin{table}[h!]
\begin{center}
\begin{tabular}{c|ccc}
Mode       & QCDF \cite{Beneke:2005pu} & pQCD \cite{Li:2006vq,Li:2005kt} & SCET \cite{Williamson:2006hb} \\
\hline
${\pi^0 K_S}$ & $0.07^{+0.05}_{-0.04}$ & $0.053^{+0.02}_{-0.03}$ & $0.077\pm0.030$\\
${\rho^0 K_S}$ & $-0.08^{+0.08}_{-0.12}$ & $-0.187^{+0.10}_{-0.06}$& $-$ \\
${\omega K_S}$ & $0.13 \pm 0.08$ & $0.153^{+0.03}_{-0.07}$& $-$ \\
\end{tabular}
\end{center}
\end{table}
}

It is interesting to note that $\Delta S_{\rho K_S}$ is the only one that is predicted to be negative, while all experimental central values are negative (see Fig. \ref{fig:other2-body}).  According to the analysis \cite{Cheng:2005bg} final state interactions could change appreciably $S_{\omega K_S}$, $S_{\rho K_S}$, but even then one still has $\Delta S_f$ $\sim O(0.1)$.

\section{Three-body modes}
In \cite{Gershon:2004tk} it was noted that $B\to \pi^0\pi^0K_S$ and  $B\to K_SK_SK_S$ are CP-even over the entire phase space so that no dilution of $S_f$ occurs in the integration over the phase space. This nice property does not hold for  $B\to K^+K^-K_S$ where both CP-even and CP-odd components are present. Nevertheless, an analysis based on isospin shows that  $B\to K^+K^-K_S$ away from $\phi K_S$ is mostly CP even \cite{Gronau:2005ax,Garmash:2003er}. 

Since there are no $b\to u\bar u s$ tree contributions in $B\to K_SK_SK_S$ one would naively expect $\Delta S_{K_SK_SK_S}$ to be very small, and for the other $\Delta S_{KKK}$ to be $\sim O(0.1)$. However, a calculation based on HM$\chi$PT, a model of form factors and a model of  non-resonant amplitude behaviour gives all $\Delta S_f\sim 0.05$ \cite{Cheng:2005ug,Cheng:2007si}. More work is needed to confirm this observation.

\section{Conclusions}
The experimental values of $\Delta S_f$ are found to be negative  in all modes and are also consistently lower than the theoretical predictions. It is a bit more difficult to assign a statistical significance to this statement, however. It is clear that different decay modes have different "tree pollutions", with $\phi K_S$ and $\eta' K_S$
being the cleanest. Simply averaging the experimental values for $\Delta S_f$ over different modes  is not correct, since the "tree pollution" is not negligible
compared to the experimental errors. To ascertain whether the experimental values of $\Delta S_f$ represent a deviation from SM or not the use of theory therefore cannot be avoided. 

The question is: how to take into account the theory? If all three approaches, QCDF, SCET and pQCD gave identical predictions, there would have been no problem. While this is not the case, the three approaches do give comparable predictions for different modes, with the difference attributable to different treatments of
higher order corrections. None of the treatments thus seems to be clearly wrong either. 

I would like to advertise two prescriptions that are both conservative and fairly intuitive. The first one is to take the theoretical framework in which the largest
number of predictions has been made and only average over modes where there are theoretical predictions, while dropping the remaining experimental results (alas!). The largest set of predictions for different modes is at present available in QCDF \cite{Beneke:2005pu}. Taking the lowest $(\Delta S_f)_{\rm Th}$ value obtained in the scan over QCDF input parameters in \cite{Beneke:2005pu} and then averaging the difference
\beq
(\Delta {S_f})_{\rm Corr}=(\Delta S_f)_{\rm Exp}-(\Delta S_f)_{\rm Th},
\eeq
by using only the experimental errors, gives
\beq
\begin{split}
\overline{(\Delta {S_f})}_{\rm Corr}&=\sin 2 \beta^{\rm Peng}-\sin 2 \beta^{\rm Tree}\\
&=-0.133\pm0.063 \quad ( > 2.1 \sigma {\rm ~effect}). \label{result}
\end{split}
\eeq
In the above average the $3-$body modes and the $f_0 K_S$ mode were dropped since there are no predictions for the corresponding $S_f$ in QCDF.
The error in Eq. \eqref{result} does not have a clear statistical meaning. Nevertheless, I believe the correct interpretation of the above result is that we have an effect that
is  larger than $0.133/0.063=2.1$ $\sigma$. 

The other conservative prescription is that for each $(\Delta {S_f})_{\rm Corr}$ one takes the smallest value predicted from the three theoretical approaches, QCDF, SCET and pQCD, and then averages over modes while adding quadratically theoretical and experimental errors. Curiously enough this gives at present almost exactly the same result as quoted for the first prescription in Eq. \eqref{result} above. 



\begin{acknowledgments}
The work of J.Z. is supported in 
part by the European Commission RTN network, Contract No.~MRTN-CT-2006-035482 (FLAVIAnet) and by the Slovenian Research Agency. 
\end{acknowledgments}

\bigskip 

\begin{thebibliography}{99}   




\bibitem{Bigi:1981qs}
  I.~I.~Y.~Bigi and A.~I.~Sanda,
  Nucl.\ Phys.\  B {\bf 193}, 85 (1981).

\bibitem{London:1989ph}
  D.~London and R.~D.~Peccei,
  Phys.\ Lett.\  B {\bf 223}, 257 (1989).



\bibitem{Grossman:1996ke}
  Y.~Grossman and M.~P.~Worah,
  Phys.\ Lett.\  B {\bf 395}, 241 (1997)
  [arXiv:hep-ph/9612269].

\bibitem{Barberio:2007cr}
  E.~Barberio {\it et al.}  [Heavy Flavor Averaging Group (HFAG)
                  Collaboration],
  arXiv:0704.3575 [hep-ex] and online update at http://www.slac.stanford.edu/xorg/hfag

\bibitem{Gronau:1989ia}
  M.~Gronau,
  Phys.\ Rev.\ Lett.\  {\bf 63}, 1451 (1989).



\bibitem{Boos:2004xp}
  H.~Boos, T.~Mannel and J.~Reuter,
  Phys.\ Rev.\  D {\bf 70}, 036006 (2004)
  [arXiv:hep-ph/0403085].

\bibitem{Ciuchini:2005mg}
  M.~Ciuchini, M.~Pierini and L.~Silvestrini,
  Phys.\ Rev.\ Lett.\  {\bf 95}, 221804 (2005)
  [arXiv:hep-ph/0507290].

\bibitem{Li:2006vq}
  H.~n.~Li and S.~Mishima,
  JHEP {\bf 0703}, 009 (2007)
  [arXiv:hep-ph/0610120].



\bibitem{Grossman:2003qp}
  Y.~Grossman, Z.~Ligeti, Y.~Nir and H.~Quinn,
  Phys.\ Rev.\  D {\bf 68}, 015004 (2003)
  [arXiv:hep-ph/0303171].

\bibitem{Gronau:2004hp}
  M.~Gronau, J.~L.~Rosner and J.~Zupan,
  Phys.\ Lett.\  B {\bf 596}, 107 (2004)
  [arXiv:hep-ph/0403287].

\bibitem{Gronau:2003kx}
  M.~Gronau, Y.~Grossman and J.~L.~Rosner,
  Phys.\ Lett.\  B {\bf 579}, 331 (2004)
  [arXiv:hep-ph/0310020].

\bibitem{Gronau:2006qh}
  M.~Gronau, J.~L.~Rosner and J.~Zupan,
  Phys.\ Rev.\  D {\bf 74}, 093003 (2006)
  [arXiv:hep-ph/0608085].

\bibitem{Gronau:2005gz}
  M.~Gronau and J.~L.~Rosner,
  Phys.\ Rev.\  D {\bf 71}, 074019 (2005)
  [arXiv:hep-ph/0503131].



\bibitem{Raz:2005hu}
  G.~Raz,
  arXiv:hep-ph/0509125.

\bibitem{Engelhard:2005ky}
  G.~Engelhard and G.~Raz,
  Phys.\ Rev.\  D {\bf 72}, 114017 (2005)
  [arXiv:hep-ph/0508046].

\bibitem{Engelhard:2005hu}
  G.~Engelhard, Y.~Nir and G.~Raz,
  Phys.\ Rev.\  D {\bf 72}, 075013 (2005)
  [arXiv:hep-ph/0505194].

\bibitem{Chiang:2003pm}
  C.~W.~Chiang, M.~Gronau, Z.~Luo, J.~L.~Rosner and D.~A.~Suprun,
  Phys.\ Rev.\  D {\bf 69}, 034001 (2004)
  [arXiv:hep-ph/0307395].

\bibitem{Chiang:2004nm}
  C.~W.~Chiang, M.~Gronau, J.~L.~Rosner and D.~A.~Suprun,
  Phys.\ Rev.\  D {\bf 70}, 034020 (2004)
  [arXiv:hep-ph/0404073].

 
  
\bibitem{Buras:2003dj}
  A.~J.~Buras, R.~Fleischer, S.~Recksiegel and F.~Schwab,
  Phys.\ Rev.\ Lett.\  {\bf 92}, 101804 (2004)
  [arXiv:hep-ph/0312259];
  Nucl.\ Phys.\  B {\bf 697}, 133 (2004)
  [arXiv:hep-ph/0402112];
  Eur.\ Phys.\ J.\  C {\bf 45}, 701 (2006)
  [arXiv:hep-ph/0512032].

\bibitem{Fleischer:2007mq}
  R.~Fleischer, S.~Recksiegel and F.~Schwab,
  Eur.\ Phys.\ J.\  C {\bf 51}, 55 (2007)
  [arXiv:hep-ph/0702275].


\bibitem{Aubert:2006wv}
  B.~Aubert {\it et al.}  [BABAR Collaboration],
  Phys.\ Rev.\ Lett.\  {\bf 98}, 031801 (2007)
  [arXiv:hep-ex/0609052].
  
\bibitem{Aubert:2006ad}
  B.~Aubert {\it et al.}  [BABAR Collaboration],
  arXiv:hep-ex/0607096.
  
\bibitem{Chen:2006nk}
  K.~F.~Chen {\it et al.}  [Belle Collaboration],
  Phys.\ Rev.\ Lett.\  {\bf 98}, 031802 (2007)
  [arXiv:hep-ex/0608039].


  
\bibitem{Hara:Belle}
K. Hara [Belle Collaboration], presented at ICHEP06, Moscow.

\bibitem{Bauer:2000ew}
  C.~W.~Bauer, S.~Fleming and M.~E.~Luke,
  Phys.\ Rev.\  D {\bf 63}, 014006 (2001)
  [arXiv:hep-ph/0005275].
  C.~W.~Bauer, S.~Fleming, D.~Pirjol and I.~W.~Stewart,
  Phys.\ Rev.\ D {\bf 63}, 114020 (2001)
  [arXiv:hep-ph/0011336];
  C.~W.~Bauer and I.~W.~Stewart,
  Phys.\ Lett.\ B {\bf 516}, 134 (2001)
  [arXiv:hep-ph/0107001].


\bibitem{Mantry:2003uz}
  S.~Mantry, D.~Pirjol and I.~W.~Stewart,
  Phys.\ Rev.\  D {\bf 68}, 114009 (2003)
  [arXiv:hep-ph/0306254].

\bibitem{Soni:2005jj}
  A.~Soni and J.~Zupan,
  Phys.\ Rev.\  D {\bf 75}, 014024 (2007)
  [arXiv:hep-ph/0510325].



\bibitem{Chay:2006ve}
  J.~Chay, C.~Kim, A.~K.~Leibovich and J.~Zupan,
  Phys.\ Rev.\  D {\bf 74}, 074022 (2006)
  [arXiv:hep-ph/0607004].

\bibitem{Beneke:1999br}
  M.~Beneke, G.~Buchalla, M.~Neubert and C.~T.~Sachrajda,
  Phys.\ Rev.\ Lett.\  {\bf 83}, 1914 (1999)
  [arXiv:hep-ph/9905312];
  Nucl.\ Phys.\ B {\bf 591}, 313 (2000)
  [arXiv:hep-ph/0006124];
  M.~Beneke, G.~Buchalla, M.~Neubert and C.~T.~Sachrajda,
  Nucl.\ Phys.\ B {\bf 606}, 245 (2001)
  [arXiv:hep-ph/0104110].
 
\bibitem{Beneke:2002jn}
  M.~Beneke and M.~Neubert,
  Nucl.\ Phys.\ B {\bf 651}, 225 (2003)
  [arXiv:hep-ph/0210085].

\bibitem{Beneke:2003zv}
  M.~Beneke and M.~Neubert,
  Nucl.\ Phys.\ B {\bf 675}, 333 (2003)
  [arXiv:hep-ph/0308039].


\bibitem{Bauer:2004tj}
  C.~W.~Bauer, D.~Pirjol, I.~Z.~Rothstein and I.~W.~Stewart,
  Phys.\ Rev.\ D {\bf 70}, 054015 (2004)
  [arXiv:hep-ph/0401188];
  C.~W.~Bauer, I.~Z.~Rothstein and I.~W.~Stewart,
  Phys.\ Rev.\  D {\bf 74}, 034010 (2006)
  [arXiv:hep-ph/0510241].

\bibitem{Jain:2007dy}
  A.~Jain, I.~Z.~Rothstein and I.~W.~Stewart,
  arXiv:0706.3399 [hep-ph].

\bibitem{Williamson:2006hb}
  A.~R.~Williamson and J.~Zupan,
  Phys.\ Rev.\  D {\bf 74}, 014003 (2006)
  [arXiv:hep-ph/0601214].


\bibitem{Bauer:2005wb}
  C.~W.~Bauer, D.~Pirjol, I.~Z.~Rothstein and I.~W.~Stewart,
  Phys.\ Rev.\  D {\bf 72}, 098502 (2005)
  [arXiv:hep-ph/0502094].

\bibitem{Beneke:2004bn}
  M.~Beneke, G.~Buchalla, M.~Neubert and C.~T.~Sachrajda,
  Phys.\ Rev.\  D {\bf 72}, 098501 (2005)
  [arXiv:hep-ph/0411171].


\bibitem{Chay:2003ju}
  J.~Chay and C.~Kim,
  Nucl.\ Phys.\  B {\bf 680}, 302 (2004)
  [arXiv:hep-ph/0301262].


\bibitem{Beneke:2005vv}
  M.~Beneke and S.~Jager,
  Nucl.\ Phys.\  B {\bf 751}, 160 (2006)
  [arXiv:hep-ph/0512351];
  M.~Beneke and S.~Jager,
  Nucl.\ Phys.\  B {\bf 768}, 51 (2007)
  [arXiv:hep-ph/0610322].

\bibitem{Bell:2007tv}
  G.~Bell,
  arXiv:0705.3127 [hep-ph].
  
\bibitem{Hill:2004if}
  R.~J.~Hill, T.~Becher, S.~J.~Lee and M.~Neubert,
  JHEP {\bf 0407}, 081 (2004)
  [arXiv:hep-ph/0404217].

\bibitem{Kirilin:2005xz}
  G.~G.~Kirilin,
  arXiv:hep-ph/0508235.

\bibitem{Beneke:2005gs}
  M.~Beneke and D.~Yang,
  Nucl.\ Phys.\  B {\bf 736}, 34 (2006)
  [arXiv:hep-ph/0508250].

\bibitem{Beneke:2005pu}
  M.~Beneke,
  Phys.\ Lett.\  B {\bf 620}, 143 (2005)
  [arXiv:hep-ph/0505075].

\bibitem{Li:2005kt}
  H.~n.~Li, S.~Mishima and A.~I.~Sanda,
  Phys.\ Rev.\  D {\bf 72}, 114005 (2005)
  [arXiv:hep-ph/0508041].

\bibitem{Li:2006jv}
  H.~n.~Li and S.~Mishima,
  Phys.\ Rev.\  D {\bf 74}, 094020 (2006)
  [arXiv:hep-ph/0608277].
  

\bibitem{Ciuchini:2001gv}
  M.~Ciuchini, E.~Franco, G.~Martinelli, M.~Pierini and L.~Silvestrini,
  Phys.\ Lett.\  B {\bf 515}, 33 (2001)
  [arXiv:hep-ph/0104126].



\bibitem{Silvestrini:2007yf}
  L.~Silvestrini,
  arXiv:0705.1624 [hep-ph].

\bibitem{Buras:1998ra}
  A.~J.~Buras and L.~Silvestrini,
  Nucl.\ Phys.\  B {\bf 569}, 3 (2000)
  [arXiv:hep-ph/9812392].


\bibitem{Cheng:2005bg}
  H.~Y.~Cheng, C.~K.~Chua and A.~Soni,
  Phys.\ Rev.\  D {\bf 72}, 014006 (2005)
  [arXiv:hep-ph/0502235].

\bibitem{Lipkin:1990us}
  H.~J.~Lipkin,
  Phys.\ Lett.\ B {\bf 254}, 247 (1991).

\bibitem{Lipkin:1998ew}
  H.~J.~Lipkin,
  Phys.\ Lett.\ B {\bf 433}, 117 (1998).

\bibitem{Gerard:2006ch}
  J.~M.~Gerard and E.~Kou,
  Phys.\ Rev.\ Lett.\  {\bf 97}, 261804 (2006)
  [arXiv:hep-ph/0609300].

\bibitem{Gershon:2004tk}
  T.~Gershon and M.~Hazumi,
  Phys.\ Lett.\  B {\bf 596}, 163 (2004)
  [arXiv:hep-ph/0402097].
  
\bibitem{Gronau:2005ax}
  M.~Gronau and J.~L.~Rosner,
  Phys.\ Rev.\  D {\bf 72}, 094031 (2005)
  [arXiv:hep-ph/0509155].

\bibitem{Garmash:2003er}
  A.~Garmash {\it et al.}  [Belle Collaboration],
  Phys.\ Rev.\  D {\bf 69}, 012001 (2004)
  [arXiv:hep-ex/0307082].



\bibitem{Cheng:2005ug}
  H.~Y.~Cheng, C.~K.~Chua and A.~Soni,
  Phys.\ Rev.\  D {\bf 72}, 094003 (2005)
  [arXiv:hep-ph/0506268].

\bibitem{Cheng:2007si}
  H.~Y.~Cheng, C.~K.~Chua and A.~Soni,
  arXiv:0704.1049 [hep-ph].



\end{thebibliography}

\end{document}